\documentclass[compsoc,conference,a4paper,10pt,times]{IEEEtran}

\usepackage{cite}
\usepackage[dvipsnames,x11names]{xcolor}
\usepackage{pifont}
\usepackage{graphicx}
\usepackage{scalefnt}
\usepackage{tikzpeople}
\usepackage{oplotsymbl}
\usepackage{fontawesome5}
\usepackage{circledsteps}
\usepackage[edges]{forest}
\usepackage{tabularx}
\usepackage{adjustbox}
\usepackage{keyvaltable}
\usepackage{threeparttable}
\usepackage{listings}
\usepackage[hyphens]{url}
\usepackage[colorlinks=true,urlcolor=black]{hyperref}
\usepackage[hyphenbreaks]{breakurl}
\usepackage[textsize=footnotesize]{todonotes}

\usetikzlibrary{patterns,arrows,arrows.meta,shapes.geometric,calc}

\ifdefined\AnonymousReview
  \setuptodonotes{disable}
\fi

\ifdefined\AnonymousReview

\fi

\definecolor{ACMRed}{rgb}{.992,.106,.078}
\definecolor{ACMOrange}{rgb}{.988,.573,0}
\definecolor{ACMPurple}{rgb}{.396,.004,.42}
\definecolor{ACMBlue}{rgb}{.004,0.51,.675}
\definecolor{ACMDarkBlue}{rgb}{.035,.208,.478}
\definecolor{KubernetesBlue}{rgb}{.196,.424,.898}

\begin{document}

\setlength{\marginparwidth}{15mm}

\makeatletter
\@ifdefinable\anon{%
  \newcommand{\anon}[2][ANONYMIZED]{%
    \@ifundefined{AnonymousReview}{#2}{#1}%
  }%
}
\makeatother



\newcommand\YAMLvaluestyle{\ttfamily\footnotesize\linespread{0.9}\color{black}\mdseries}
\newcommand\YAMLkeystyle{\ttfamily\footnotesize\linespread{0.9}\color{ACMDarkBlue}\bfseries}
\newcommand\YAMLkeywordstyle{\ttfamily\footnotesize\linespread{0.9}\color{darkgray}\bfseries}
\newcommand\YAMLcolonstyle{\ttfamily\footnotesize\linespread{0.9}\color{ACMDarkBlue}\mdseries}
\newcommand\YAMLcommentstyle{\ttfamily\footnotesize\linespread{0.9}\color{ACMPurple}\mdseries}
\lstdefinelanguage{yaml}{
  keywords={true,false,null,y,n},
  keywordstyle=\YAMLkeywordstyle,
  basicstyle=\YAMLkeystyle,
  sensitive=false,
  comment=[l]{\#},
  morecomment=[s]{/*}{*/},
  commentstyle=\YAMLcommentstyle,
  stringstyle=\YAMLvaluestyle,
  moredelim=[l][\color{ACMOrange}]{\&},
  moredelim=[l][\color{ACMPurple}]{*},
  moredelim=**[il][\YAMLcolonstyle{:}\YAMLvaluestyle]{:},
  morestring=[b]',
  morestring=[b]",
  rulecolor=\color{black},
  escapeinside={(*@}{@*)},
}

\newcommand{\koney}{Koney}
\newcommand{\koneyAllCaps}{KONEY}
\newcommand{\repository}{https://github.com/dynatrace-oss/koney}
\ifdefined\AnonymousReview
  \renewcommand{\koney}{BeeOp}
  \renewcommand{\koneyAllCaps}{BeeOp}
  \renewcommand{\repository}{https://github.com}
  \IEEEspecialpapernotice{``\koney{}'' is an anonymized name.
    The final name will be provided after review.}
\else\fi

\title{\koney{}: A Cyber Deception Orchestration Framework for Kubernetes}

\ifdefined\AnonymousReview
  \author{\IEEEauthorblockN{Anonymous Author(s)}
    \IEEEauthorblockA{\ignorespaces}
    \IEEEauthorblockA{\ignorespaces}
  }
\else
  \author{%
    \IEEEauthorblockN{%
      \href{https://orcid.org/0000-0002-6820-4953}{Mario Kahlhofer}}
    \IEEEauthorblockA{%
      Dynatrace Research}
    \IEEEauthorblockA{%
      mario.kahlhofer@dynatrace.com
    }
    \and
    \IEEEauthorblockN{%
      \href{https://orcid.org/0000-0002-8743-0825}{Matteo Golinelli}}
    \IEEEauthorblockA{%
      University of Trento}
    \IEEEauthorblockA{%
      matteo.golinelli@unitn.it
    }
    \and
    \IEEEauthorblockN{%
      \href{https://orcid.org/0000-0003-2821-2489}{Stefan Rass}}
    \IEEEauthorblockA{%
      Johannes Kepler University Linz}
    \IEEEauthorblockA{%
      stefan.rass@jku.at
    }
  }
\fi

\lstset{%
  tabsize=2,
  frame=single,
  emptylines=1,
  captionpos=t,
  aboveskip=8pt,
  belowskip=12pt,
  showspaces=false,
  keepspaces=true,
  columns=fullflexible,
  showstringspaces=false,
  escapeinside={(*@}{@*)},
  basicstyle=\linespread{0.9}\footnotesize\ttfamily
}

\makeatletter
\def\lst@makecaption{%
  \def\@captype{table}%
  \@makecaption
}
\makeatother

\newcommand{\citeCloudBasedHoneypots}{\cite{%
    Memari2014:VirtualHoneynetBased, 
    Kedrowitsch2017:FirstLookUsing, 
    Osman2019:SandnetHighQuality, 
    Gupta2021:HoneyKubeDesigningHoneypot, 
    Gupta2023:HoneyKubeDesigningDeploying, 
    Machmeier2023:HoneypotImplementationCloud, 
    Spahn2023:ContainerOrchestrationHoneypot, 
    Reti2021:EscapeFakeIntroducing, 
    Li2022:OptimalDefensiveDeception, 
    Priya2023:ContainerizedCloudbasedHoneypot, 
    Zambianco2024:ResourceAwareCyberDeception, 
    Santoro2024:DemoCloudnativeCyber, 
    TheKubernetesStormCenterAuthors:KubernetesStormCenter 
  }}

\newcommand{\citeReverseProxySolutions}{\cite{%
    Han2017:EvaluationDeceptionBasedWeb,
    Araujo2014:PatchesHoneyPatchesLightweight,
    Barron2021:ClickThisNot,
    Fraunholz2018:CloxyContextawareDeceptionasaService,
    Sahin2020:LessonsLearnedSunDEW,
    Pohl2015:BHiveZeroConfiguration
  }}

\newcommand{\citeGenerativeHoneypots}{\cite{%
    McKee2023:ChatbotsHoneypotWorld, 
    Sladic2024:LLMShellGenerative, 
    Reti2024:ActHoneytokenGenerator, 
    Nguyen2021:HoneyCodeAutomatingDeceptive, 
    Karimi:GalahLLMpoweredWeb, 
    Splunk:DECEIVEDECeptionEvaluative 
  }}

\newcommand{\citeSupportedUseCases}{\cite{%
    Spitzner2003:HoneytokensOtherHoneypot, 
    Yuill2004:HoneyfilesDeceptiveFiles, 
    BenSalem2011:DecoyDocumentDeployment, 
    Voris2015:FoxTrapThwarting, 
    Angeli2024:FalseFlavorHoneypot, 
    Fraunholz2018:DefendingWebServers, 
    Fraunholz2018:CloxyContextawareDeceptionasaService, 
    Han2017:EvaluationDeceptionBasedWeb, 
    Gavrilis2007:FlashCrowdDetection, 
    Kahlhofer2024:HoneyquestRapidlyMeasuring, 
    Sahin2020:LessonsLearnedSunDEW, 
    Petrunic2015:HoneytokensActiveDefense, 
    Barron2021:ClickThisNot 
  }}

\newcommand{\numExecOne}{\Circled[fill color=black, inner color=white, outer color=white]{1}}
\newcommand{\numExecTwo}{\Circled[fill color=KubernetesBlue, inner color=white, outer color=white]{2A}}
\newcommand{\numExecTwoAlt}{\Circled[fill color=KubernetesBlue, inner color=white, outer color=white]{2B}}
\newcommand{\numExecThree}{\Circled[fill color=ACMPurple, inner color=white, outer color=white]{3}}
\newcommand{\numExecFour}{\Circled[fill color=ACMPurple, inner color=white, outer color=white]{4}}

\newcommand{\numIstioOne}{\Circled[fill color=black, inner color=white, outer color=white]{1}}
\newcommand{\numIstioTwo}{\Circled[fill color=ACMPurple!50!ACMRed, inner color=white, outer color=white]{2}}
\newcommand{\numIstioThree}{\Circled[fill color=ACMPurple!50!ACMRed, inner color=white, outer color=white]{3}}

\maketitle

\begin{abstract}
  System operators responsible for protecting software applications
  remain hesitant to implement cyber deception technology,
  including methods that place traps to catch attackers,
  despite its proven benefits. 
  Overcoming their concerns removes a barrier that currently hinders
  industry adoption of deception technology.
  Our work introduces deception policy documents to describe deception technology ``as code''
  and pairs them with \koney{}, a Kubernetes operator, which facilitates
  the setup, rotation, monitoring, and removal of traps in Kubernetes.
  We leverage cloud-native technologies, such as service meshes and eBPF, 
  to automatically add traps to containerized software applications,
  without having access to the source code.
  We focus specifically on operational properties,
  such as maintainability, scalability, and simplicity,
  which we consider essential to accelerate the adoption of cyber deception technology
  and to facilitate further research on cyber deception. 
\end{abstract}


\section{Introduction}
\label{sec:introduction}

The interest in cyber deception technology,
which includes methods that set subtle traps to easily spot,
delay, and deter attackers~\cite{Ferguson-Walter2023:CyberExpertFeedback},
has recently gained renewed attention.
The emergence of generative honeypots~\citeGenerativeHoneypots{}
now allows the rapid generation of enticing deceptive payloads.
Simultaneously, modern software application architectures,
particularly cloud-native architectures, have become increasingly dynamic and flexible,
offering novel opportunities to integrate active defensive measures%
~\cite{Kahlhofer2024:ApplicationLayerCyber}.
We believe that leveraging these opportunities
can accelerate the adoption of cyber deception technology by the industry,
as they can address two common, yet unexplored, needs of system operators:

\begin{enumerate}
  \item \textbf{Policy Documents.}
        How can cyber deception techniques be formalized ``as code'',
        i.e., expressed as structured documents?
  \item \textbf{Automated Deployment.}
        How can we implement these policies in real-world software systems,
        without the ability to own or alter the application's source code,
        which is a common constraint faced by system operators?
\end{enumerate}

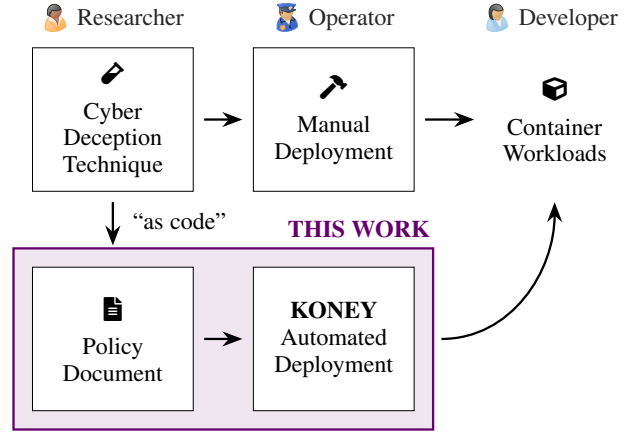
\begin{figure}[t]
  \centering
  \resizebox{\columnwidth}{!}{{\scalefont{1.25}\begin{tikzpicture}[scale=1.0, auto, decoration={snake, amplitude=6pt, segment length=1.5mm, pre length=1pt, post length=6pt}]
  \pgfdeclarelayer{background}
  \pgfdeclarelayer{people}
  \pgfsetlayers{background,people,main}

  \tikzstyle{block} = [rectangle, draw, text width=6em, text centered, minimum height=6em, node distance=1cm and 0.7cm, outer sep=0.2cm]
  \tikzstyle{line} = [draw, -{Stealth[scale=1.25]}, very thick]

  \node [block] (papers) {\faVial\\\vspace{0.4em}Cyber Deception Technique};
  \node [block, right=of papers] (manual) {\faHammer\\\vspace{0.4em}Manual Deployment};
  \node [block, draw=none, right=of manual] (apps) {\faCube\\\vspace{0.4em}Container Workloads};

  \node [block, below=of papers, fill=white] (policy) {\faFile*\\\vspace{0.4em}Policy Document};
  \node [block, below=of manual, fill=white] (koney) {\textbf{\koneyAllCaps{}}\\Automated Deployment};

  \begin{pgfonlayer}{background}
    \node [block, draw=ACMPurple, fill=ACMPurple!10, very thick, fit=(policy) (koney)] (focus) {};
    \node [above left, shift={(-0.1cm,-0.15cm)}, text=ACMPurple] at (focus.north east) {\bfseries THIS WORK};
  \end{pgfonlayer}

  \begin{pgfonlayer}{people}
    \node [alice, node distance=0.1cm, scale=1, above=of papers, shift={(-1.05cm,0)}] (alice) {};
    \node [above right, anchor=west, text=black!85] at (alice.10) {Researcher};
    \node [police, node distance=0.1cm, scale=1, above=of manual, shift={(-0.85cm,0)}] (bob) {};
    \node [above right, anchor=west, text=black!85, shift={(0,-0.05cm)}] at (bob.10) {Operator};
    \node [dave, female, node distance=0.1cm, scale=1, above=of apps, shift={(-1.05cm,0)}] (charlie) {};
    \node [above right, anchor=west, text=black!85, shift={(0,-0.05cm)}] at (charlie.10) {Developer};
  \end{pgfonlayer}

  \path [line] (papers) -- node [align=left, shift={(0.2cm,0.05cm)}] {``as code''} ($(policy.north)+(0,0.2cm)$);
  \path [line] (papers) -- node [align=center, shift={(0,0.1cm)}] {} (manual);
  \path [line] (manual) -- ($(apps.west)+(0.2cm,0)$);
  \path [line] (policy) -- (koney);
  \path [line] ($(koney.east)+(0.3cm,0)$) to[out=0,in=-90] (apps);
\end{tikzpicture}}}
  \caption{State of the art in deploying cyber deception and this work.}
  \label{fig:intro}
\end{figure}

This offers system operators an alternative
to deciphering cyber deception methods from academic papers
and subsequently tailoring their understanding to fit a particular technical environment
(Figure~\ref{fig:intro}).


We introduce \koney{} as a tool for the automated deployment of cyber deception policies.
While \koney{} is designed for Kubernetes, 
the design decisions we present can be leveraged
by cloud- and container-based deception research~\citeCloudBasedHoneypots{}.
As the deployment of classic honeypot software is well-explored%
~\cite{
  Fraunholz2017:DeploymentStrategiesDeception,
  Nawrocki2016:SurveyHoneypotSoftware},
we initially concentrate on techniques close to the application layer.

Our main contributions are as follows:
\begin{enumerate}
  \item A schema for describing \textbf{cyber deception policies},
        such as honeytokens in file systems and traps that protect HTTP-based web applications,
        but not limited to application layer techniques.
  \item A framework to automate the setup, rotation, monitoring, and removal
        of cyber deception technology in container orchestration platforms.
  \item \textbf{\koney{}}, an open-source%
        \footnote{\anon[Repository URL anonymized. Artifact provided after review.]{\url{\repository}}}
        operator that automates application layer cyber deception in Kubernetes.
\end{enumerate}

\section{Problem Statement}
\label{sec:problem}


Studies note that ``organizations are still reluctant to implement cyber deception''%
~\cite{Tounsi2022:CyberDeceptionUltimate}.
One potential obstacle for system operators could be the technical challenges
that researchers have also previously encountered%
~\cite{
  Han2018:DeceptionTechniquesComputer,
  Sahin2020:LessonsLearnedSunDEW}.
Kahlhofer and Rass 
note that ``a barrier to widespread adoption of cyber deception technology
is its deployment in real-world software systems''%
~\cite{Kahlhofer2024:ApplicationLayerCyber}.
They suggest necessary properties that such technologies must have,
including operational aspects like simplicity, maintainability, and scalability,
and technical aspects such as being resource-efficient,
the ability to transparently modify or redirect application data,
and avoiding application restarts.
We designed \koney{} with these properties in mind.

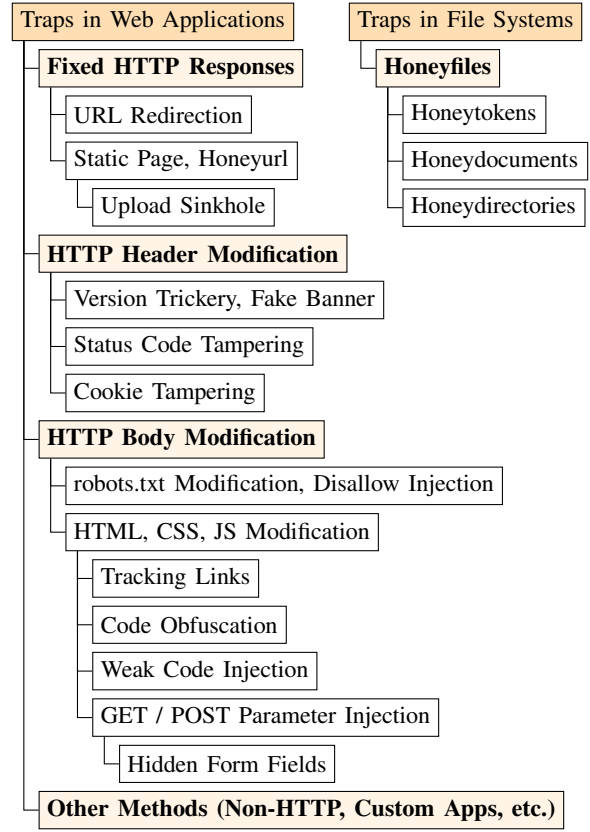
\begin{figure}[t]
  \centering
  {\small\forestset{title/.style={fill=ACMOrange!30}}
\forestset{head/.style={fill=ACMOrange!10,font=\bfseries}}
\forestset{ok/.style={fill=YellowGreen!30}}
\forestset{leaf/.style={draw=none,no edge,color=gray,align=left,font=\footnotesize}}

\begin{forest}
  for tree={%
  draw,
  minimum height=14pt,
  s sep=3pt,
  },
  where level=0{%
  s sep=20pt
  }{%
  folder,
  grow'=0,
  if level=1{%
  before typesetting nodes={child anchor=north},
  edge path'={(!u.parent anchor) -- ++(0,-5pt) -| (.child anchor)},
  }{}
  }
  [,phantom
  [Traps in Web Applications,title
  [Fixed HTTP Responses,head
  [URL Redirection]
  [{Static Page, Honeyurl}
    [Upload Sinkhole]]
  ]
  [HTTP Header Modification,head
  [{Version Trickery, Fake Banner}]
  [Status Code Tampering]
  [Cookie Tampering]
  ]
  [HTTP Body Modification,head
  [{robots.txt Modification, Disallow Injection}]
  [{HTML, CSS, JS Modification}
    [Tracking Links]
    [Code Obfuscation]
    [Weak Code Injection]
    [GET / POST Parameter Injection
        [Hidden Form Fields]]
  ]
  ]
  [{Other Methods (Non-HTTP, Custom Apps, etc.)},head]
  ]
  [Traps in File Systems,title
  [Honeyfiles,head
  [Honeytokens]
  [Honeydocuments]
  [Honeydirectories]
  ]
  ]
  ]
\end{forest}}
  \caption{Example use cases of application layer cyber deception.}
  \label{fig:usecases}
\end{figure}

Ultimately, we aim to deploy many
previously published cyber deception techniques%
~\citeSupportedUseCases{}
with \koney{}.
Han~et~al. 
categorize techniques into network, system, application, and data layers%
~\cite{Han2018:DeceptionTechniquesComputer}.
Deploying classic honeypots is typically simple since they operate on the network layer,
where additional workloads can be easily connected,
while application and system layer techniques require more pervasive technical methods%
~\cite{Kahlhofer2024:ApplicationLayerCyber}.
Our work primarily focuses on the application layer,
for which we identified a representative sample of use cases (Figure~\ref{fig:usecases}).
We will describe these use cases with policy documents 
and deploy them to a Kubernetes cluster using \koney{}.


\textbf{Honeyfiles.}
Placing fake files, security tokens, or documents
that appear sensitive and appealing to adversaries
is one of the original use cases of cyber deception.
We distinguish between honeytokens%
~\cite{Spitzner2003:HoneytokensOtherHoneypot}
and honeydocuments%
\cite{
  BenSalem2011:DecoyDocumentDeployment,
  Voris2015:FoxTrapThwarting,
  Yuill2004:HoneyfilesDeceptiveFiles}.
Honeytokens are small files with only a few lines of seemingly sensitive text,
whereas honeydocuments are entire files, such as Office and PDF documents.
Finally, honeydirectories are entire folder structures that contain deceptive files.

\textbf{Fixed HTTP responses.}
By injecting new, previously non-existent, HTTP endpoints into web applications,
we can lure attackers and vulnerability scanners to misleading paths%
~\cite{Angeli2024:FalseFlavorHoneypot,Fraunholz2018:DefendingWebServers}.
Typically, the aim is to either redirect HTTP requests to honeypots,
or directly respond to requests with a deceptive file or payload,
such as a fake admin page when someone probes ``/wp-admin''.
Pages that allow file uploads are called upload sinkholes%
~\cite{Fraunholz2018:CloxyContextawareDeceptionasaService}.

\textbf{HTTP header modification.}
Vulnerability scanners typically inspect only certain headers%
~\cite{Fraunholz2018:DefendingWebServers}.
To steer scanners towards incorrect assessments, we can manipulate version numbers
(e.g., in the ``Server'' header field),
send incorrect service banners to combat banner grabbing%
~\cite{Albanese2015:DeceptionBasedApproach},
tamper with session cookies to entice adversaries%
~\cite{
  Han2017:EvaluationDeceptionBasedWeb,
  Fraunholz2018:CloxyContextawareDeceptionasaService},
and alter response status codes%
~\cite{Kern2024:InjectingSharedLibraries}.

\textbf{HTTP body modification.}
To realize more invasive use cases, we need to modify the HTTP response body directly.
For example, the ``robots.txt'' file is a common entry point for scanners and adversaries
to discover routes that the site owner wants to hide from them.
Adding new routes to this file, a process known as ``disallow injection''%
~\cite{
  Han2017:EvaluationDeceptionBasedWeb,
  Fraunholz2018:CloxyContextawareDeceptionasaService,
  Fraunholz2018:DefendingWebServers},
is a valuable cyber deception technique.
Deceptive elements may also be directly added to HTML, CSS, and JavaScript sources.
For instance, we can add tracking links to register access attempts
to pages that only adversaries are likely to visit%
~\cite{Gavrilis2007:FlashCrowdDetection},
obfuscate source code by adding deceptive elements to it%
~\cite{Fraunholz2018:CloxyContextawareDeceptionasaService},
imitate real security weaknesses and vulnerabilities%
~\cite{
  Kahlhofer2024:HoneyquestRapidlyMeasuring,
  Sahin2020:LessonsLearnedSunDEW},
add deceptive GET or POST parameters such as ``admin=false'' to hyperlinks%
~\cite{
  Han2017:EvaluationDeceptionBasedWeb,
  Petrunic2015:HoneytokensActiveDefense,
  Kahlhofer2024:HoneyquestRapidlyMeasuring},
or inject hidden form fields%
~\cite{Barron2021:ClickThisNot}.

%


While we don't consider it part of our primary problem statement,
in \S\ref{sec:discussion:extensions} we describe how \koney{}
can be extended to support deception techniques for protocols other than HTTP%
~\cite{Nawrocki2016:SurveyHoneypotSoftware}
and domain-specific classes of applications, such as database systems%
~\cite{Storey2009:CatchingFliesHoney}
and WordPress plugins%
~\cite{Reti2023:SCANTRAPProtectingContent}.


\section{Kubernetes Terminology}
\label{sec:kubernetes}

This section provides an overview of key terminology and concepts related to Kubernetes.

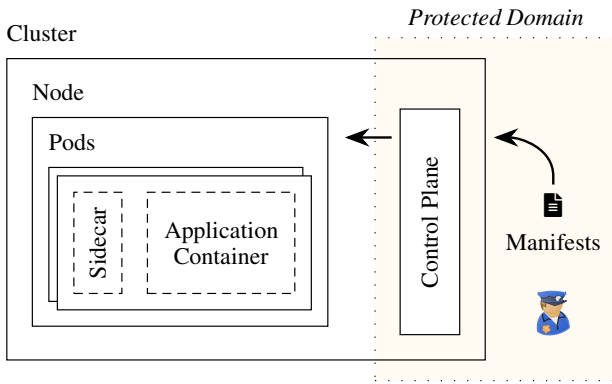
\begin{figure}[ht]
  \centering
  \resizebox{\columnwidth}{!}{{\scalefont{0.8}\begin{tikzpicture}[scale=1.0,auto]
  \pgfdeclarelayer{background}
  \pgfdeclarelayer{cluster layer}
  \pgfdeclarelayer{node layer}
  \pgfdeclarelayer{pod layer back}
  \pgfdeclarelayer{pod layer front}
  \pgfdeclarelayer{container layer}
  \pgfdeclarelayer{arrow layer}
  \pgfsetlayers{background,cluster layer,node layer,pod layer back,pod layer front,container layer,arrow layer,main}

  \tikzstyle{block} = [rectangle, draw, node distance=0.1cm, outer sep=0.1cm, align=center]
  \tikzstyle{line} = [draw, -{Stealth[scale=1.25]}, thick]

  \begin{pgfonlayer}{container layer}
    \node [block, densely dashed, inner sep=0.2cm, minimum height=4.25em, minimum width=2em] (ctr1) {};
    \node [block, densely dashed, inner sep=0.2cm, minimum height=4.25em, right=of ctr1] (ctr2) {Application\\Container}; 
    \node [anchor=center, rotate=90] at (ctr1.center) {Sidecar};
  \end{pgfonlayer}

  \begin{pgfonlayer}{pod layer front}
    \node [block, fit=(ctr1) (ctr2), fill=white] (pod1) {};
  \end{pgfonlayer}

  \begin{pgfonlayer}{pod layer back}
    \node [block, fit=(ctr1) (ctr2), fill=white, shift={(-0.1cm,0.1cm)}] (pod2) {};
    \node [above right, name=pod2-label] at (pod2.north west) {Pods};
  \end{pgfonlayer}

  \begin{pgfonlayer}{node layer}
    \node [block, fit=(pod1) (pod2) (pod2-label)] (node) {};
    \node [above right, name=node-label] at (node.north west) {Node};

    \node [block, inner sep=0, fit=(node.north east) (node.south east),
      shift={(1.1cm,0)}, minimum width=0.7cm, fill=white] (ctrl) {};
    \node [anchor=center, rotate=90] at (ctrl.center) {Control Plane};

    \node [anchor=center, shift={(1cm,0.2cm)}, name=manifests-logo] at (ctrl.east) {\faFile*};
    \node [below,name=manifests-label] at (manifests-logo.south) {Manifests};

    \node [police, node distance=0.4cm, scale=1, below=of manifests-label] (bob) {};
  \end{pgfonlayer}

  \begin{pgfonlayer}{node layer}
    \node [block, fit=(node) (node-label) (ctrl), inner sep=0.2cm] (cluster) {};
    \node [above right, name=cluster-label] at (cluster.north west) {Cluster};
  \end{pgfonlayer}

  \begin{pgfonlayer}{background}
    \node [name=ctrl-top] at ($(ctrl.north west)+(0,0.55cm)$) {};
    \node [name=ctrl-bot] at ($(ctrl.south west)-(0,0.25cm)$) {};
    \node [block, loosely dotted, fill=ACMOrange!5, fit=(ctrl-top) (ctrl-bot) (manifests-label) (manifests-logo)] (protected) {};
    \node [above, shift={(0,-0.05cm)}] at (protected.north) {\itshape Protected Domain};
  \end{pgfonlayer}

  \begin{pgfonlayer}{arrow layer}
    \path [line] ($(ctrl.west)+(0,1cm)$) -- node [below, shift={(0,-0.2cm)}] {} ($(node.east)+(0.1cm,1cm)$);
    \path [line] (manifests-logo.north) -- ($(manifests-logo.north)+(0,0.1cm)$) to[out=90,in=0] ($(ctrl.east)+(0.3cm,1cm)$);
  \end{pgfonlayer}

\end{tikzpicture}}}
  \caption{Components of a Kubernetes cluster.}
  \label{fig:kubernetes}
\end{figure}

Kubernetes is an enterprise-ready system for orchestrating container deployments.
Figure~\ref{fig:kubernetes} presents its main components.
A \emph{pod} is the smallest deployable unit and contains the software application.
A pod consists of at least one \emph{container},
and containers within the same pod share storage and network resources.
Containers used for auxiliary tasks such as monitoring are known as \emph{sidecars}.
Pods are deployed on a \emph{node}, which can be a physical or virtual machine.
A \emph{cluster} is composed of one or more nodes.
Clusters have at least one \emph{control plane},
which offers APIs to system operators to manage cluster configuration and workloads.
The control plane is typically isolated from the (application) workloads
and cannot be easily compromised.
\emph{Workloads}, such as pods, are represented by \emph{manifests},
which are formal documents that describe them. 
Manifests hold metadata, such as the resource name,
resource kind (indicating their type),
their associated \emph{namespace} (to organize resources),
\emph{labels} and \emph{annotations} (custom key-value pairs),
and their kind-specific properties.
For pods, these include container images, command-line arguments,
environment variables, mounted volumes, exposed ports, and more.

Kubernetes can be extended with \emph{operators},
which add custom logic to the Kubernetes API,
to facilitate the management of complex applications and clusters.
Operators typically install \emph{custom resource definitions (CRDs)},
which represent new kinds of resources with which system operators can interface.
\emph{Custom resources}, and more generally all resource definitions in Kubernetes,
describe a desired state.
Operators watch for changes to their custom resources (and possibly to the cluster state)
and take action to restore and maintain the desired cluster state%
~\cite{Ibryam2019:KubernetesPatterns}.
The process of restoring the state of the cluster to the desired one
is called \emph{reconciliation} and is executed in a loop by the operator,
triggered by the control plane.

\section{Cyber Deception Policies}
\label{sec:policies}

Introducing cyber deception policies enables dividing
responsibilities between three roles, which typically do not overlap:
authors of cyber deception techniques, software application developers,
and system operators deploying them. 
This pattern is especially prevalent in Kubernetes%
~\cite{Ibryam2019:KubernetesPatterns}
and is followed by popular tools to manage security policies, such as Kyverno%
~\cite{TheKyvernoAuthors:Kyverno}.
Listing~\ref{lst:boilerplate} presents the design of a policy document
and its structure in pseudo-code. 
A policy contains:

\begin{enumerate}
  \item A list of traps and their parameterization
        (\S\ref{sec:policies:fs}, \S\ref{sec:policies:web}).
        The parameters include the strategy to technically deploy the trap.
        Deployment strategies are extensively described in \S\ref{sec:operator},
        as they are implementation-specific.
        A \textbf{trap} consists of the \textbf{decoy},
        which is the entity that is attacked, 
        and the \textbf{captor}, which monitors the decoy~\cite{Fan2018:EnablingAnatomicView}.
  \item The criteria for selecting the applications and workloads
        in which to deploy the traps (\S\ref{sec:policies:select}).
\end{enumerate}

\begin{lstlisting}[
  language=yaml,
  label=lst:boilerplate,
  caption={Pseudo policy to illustrate its basic structure.}
]
kind: DeceptionPolicy
metadata:
  name: sample-boilerplate
spec:
  traps:
    - trapKindOne:      # trap kind and its
        foo: bar        # distinctive properties

      match:            # workload selection
        ...             # see (*@\YAMLcommentstyle\S\ref{sec:policies:select}@*)

      decoyDeployment:  # method for deploying
        strategy: ...   # the trap itself

      captorDeployment: # method for monitoring
        strategy: ...   # access attempts

    - trapKindOne: ...  # more traps
    - trapKindTwo: ...  # (duplicates allowed)
\end{lstlisting}

\subsection{Traps in File Systems}
\label{sec:policies:fs}

The deployment of a honeytoken in a file system
can be formally described by its \texttt{path} and \texttt{content}.
To closely resemble real security tokens, one may set the file's access mode to \texttt{readOnly}.
If the file content is longer than a few lines or is binary,
the policy must include a URL pointing to the \texttt{source} of the honeyfile
to be downloaded before deployment.
An example of both types of traps is illustrated in Listing~\ref{lst:honeyfiles}.
These two specifications can also be nested to facilitate the creation of entire honeydirectories,
as shown in Appendix~\ref{sec:appendix:examples:fs}.

\begin{lstlisting}[
  language=yaml,
  label=lst:honeyfiles,
  caption={Specification of honeytokens and -documents.}
]
- filesystemHoneytoken:
    path: /run/secrets/service_token
    content: very-secret-token
    readOnly: true
- filesystemHoneydocument:
    path: /root/passwords.docx
    source: https://srv.test/honey.docx
\end{lstlisting}

Appendix~\ref{sec:appendix:examples:fs} contains full-length sample policies
for traps in file systems. Those also illustrate two strategies for the deployment of decoys,
the ``exec'' method (\S\ref{sec:operator:container-exec})
and volume mounts (\S\ref{sec:operator:volume-mounts});
as well as the Tetragon strategy for monitoring
file access attempts (\S\ref{sec:operator:tetragon}).

\subsection{Traps in Web Applications}
\label{sec:policies:web}

To define fixed HTTP responses, we chose a formulation similar to
Baitroute~\cite{Sen:Baitroute} and HASH~\cite{DataDog:HASHHTTPAgnostic},
comprising two parts:
an expression to \texttt{match} HTTP requests by path and \texttt{method},
and the corresponding response.
Listing~\ref{lst:httpresponse} illustrates responding with predefined content, 
which resembles a trivial honeypot. This also enables responding with an HTTP redirect,
as demonstrated in Appendix~\ref{sec:appendix:examples:web}.

\begin{lstlisting}[
  language=yaml,
  label=lst:httpresponse,
  caption={Specification of a static HTTP response.}
]
- httpResponse:
    request:
      match: ^/wp-admin  # regex: starts with
      method: GET        # filter on GET only
    response:
      status: 200
      headers:
        Content-Type: text/html
      body: "<html><!-- --></html>"
\end{lstlisting}

To modify or add headers and change the status code of genuine HTTP responses,
the response specification is altered to include
a \texttt{status} and a \texttt{setHeader} directive,
as shown in Listing~\ref{lst:headermodification}.
The \texttt{removeHeaders} directive enables removing response headers from HTTP responses.

\begin{lstlisting}[
  language=yaml,
  label=lst:headermodification,
  caption={Specification for mutating HTTP headers.}
]
- httpHeaderMutation:
    request:
      match: "*"   # regex: match all paths
      method: GET  # filter on GET only
    response:
      status: 404  # override
      setHeaders:
        Server: nginx/1.2.4
        X-ApiServer: honeypot.test
      removeHeaders:
        - Cookie
\end{lstlisting}

To enable modification of the body of HTTP responses, we employ different ``engines''.
Listing~\ref{lst:bodymodification} illustrates how to use the ``regex'' engine,
which searches for a matching pattern and replaces it with another.
This engine also supports group expressions.
For example, to inject a script in the head of an HTML page
we have a group that matches the prefix plus the inner HTML \texttt{(<head>.*)}
and a group that matches the suffix \texttt{(</head>)}.
This is replaced with the first group (\texttt{\$1}),
the injection content, and the second group (\texttt{\$2}).
Note that all the examples follow the syntax of the Golang regexp package%
~\cite{TheGoAuthors:RegexpPackage}.
To further filter what responses are modified (e.g., transform HTML but not images),
we support filters in the response section, such as \texttt{matchHeaders}.

\begin{lstlisting}[
  language=yaml,
  label=lst:bodymodification,
  caption={Specification for mutating the HTTP body.}
]
- httpBodyMutation:
    request:
      match: "*"   # match all paths
      method: GET  # GET only
    response:
      matchHeaders:
        Content-Type: text/html
      bodyMutations:
        - engine: regex
          match: "(?si)(<head>.*)(</head>)"
          replace: "$1<script></script>$2"
\end{lstlisting}

Appendix~\ref{sec:appendix:examples:web} presents full-length sample policies
for traps in web applications. Decoy and captor deployment with Istio is explained in
\S\ref{sec:operator:istio-decoy} and \S\ref{sec:operator:istio-captor}, respectively.

\subsection{Selecting Resources}
\label{sec:policies:select}

Every trap needs to specify the resources it targets. \koney{} uses labels and selectors,
which is a standard method of organizing and selecting resources in Kubernetes%
~\cite{CNCF:KubernetesAPI}.
We adopt Kyverno's syntax%
~\cite{TheKyvernoAuthors:Kyverno},
because we find it well-organized and easy to use.
Moreover, since Kyverno is a popular tool,
many users might already be familiar with its syntax.
Resource selection (Listing~\ref{lst:match}) works by defining one or more ``resource filters'',
combined by \texttt{any} or \texttt{all} connectives.
\texttt{any} applies traps if any filter matches (logical OR), whereas
\texttt{all} only applies traps to resources matching every filter (logical AND).
A resource filter contains one or more distinct selectors: 

\begin{itemize}
  \item \texttt{selector.matchLabels}:
        Matches resources with exactly these labels. 
        If multiple pairs are given, resources must possess \textbf{all} of them.
  \item \texttt{selector.matchExpressions}:
        A method to formulate complex filter expressions,
        as specified in the official Kubernetes API~\cite{CNCF:KubernetesAPI}.
  \item \texttt{namespaces:}
        Matches resources within \textbf{any} namespace listed.
        If omitted, it matches all possible namespaces.
  \item \texttt{containerSelector}:
        Traps such as honeytokens operate on an individual container file system.
        This selector filters which containers are affected.
        If omitted, all containers will be affected
        (equivalent to the wildcard selector ``*'').
  \item \texttt{ports}:
        Traps such as HTTP-based traps intercept network traffic. 
        This selector filters which ports are intercepted.
        If omitted, it will match all ports.
\end{itemize}

One of \texttt{selector} or \texttt{namespaces} (or both) must be specified
to make a valid filter. 
Multiple selectors within one filter are evaluated with a logical AND. 

\begin{lstlisting}[
  language=yaml,
  label=lst:match,
  caption={Example on selecting resources.}
]
match:
  any:
    - resources:  # resource filter
        selector:
          matchLabels:
            example-label: true
        namespaces:
          - production
        containerSelector: "*"
        ports:
          - 80
\end{lstlisting}

\section{The \koney{} Operator}
\label{sec:operator}

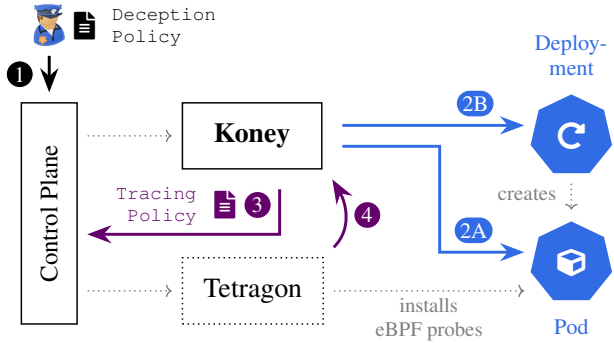
\begin{figure}[t]
  \centering
  \resizebox{\columnwidth}{!}{{\scalefont{0.75}\begin{tikzpicture}[scale=1.0,auto]
  \tikzstyle{block} = [rectangle, draw, node distance=0.6cm, outer sep=0.1cm, align=center]
  \tikzstyle{heptagon} = [regular polygon, draw, regular polygon sides=7, draw=none, fill=KubernetesBlue, text=white, outer sep=0.1cm]
  \tikzstyle{line} = [draw, -{Stealth[scale=1.0]}, thick]
  \tikzstyle{grayline} = [draw, densely dotted, -{>[scale=0.66]}, draw=gray, thin]

  \node [block, inner sep=0, minimum height=8em, minimum width=2em] (ctrl) {};
  \node [anchor=center, rotate=90, scale=0.9] at (ctrl.center) {Control Plane};

  \node [block, inner sep=0.2cm, minimum width=5em,
    right=of ctrl.north east, anchor=north west, shift={(0.2cm,0)}] (koney) {\bfseries\koney{}};
  \node [block, inner sep=0.2cm, minimum width=5em, densely dotted,
    right=of ctrl.south east, anchor=south west, shift={(0.2cm,0)}] (tetragon) {Tetragon};

  \node[heptagon, minimum size=3em, right=of koney, shift={(0.8cm,0)}] (deployment) {\faRedo*};
  \node[heptagon, minimum size=3em, below=of deployment, shift={(0,0.75cm)}] (pod) {\faCube};
  \node [above, name=deployment-label, text=KubernetesBlue, scale=0.8, align=center] at (deployment.north) {Deploy-\\ment};
  \node [below, name=pod-label, text=KubernetesBlue, scale=0.8] at (pod.south) {Pod};

  \node [police, scale=0.75, anchor=south, name=deception-policy-person, shift={(0,0.6cm)}] at (ctrl.north) {};
  \node [anchor=west, name=deception-policy-logo, right=of deception-policy-person, shift={(-1cm,0)}] {\faFile*};
  \node [below, anchor=west, align=left, name=deception-policy-label, scale=0.7]
  at (deception-policy-logo.east) {\ttfamily Deception\\\ttfamily Policy};

  \path [line] ($(deception-policy-person.south)+(0,-0.1cm)$) -- node [left, shift={(-0.07cm,0)}, scale=0.8] {\numExecOne{}} (ctrl.north);
  \path [grayline] (ctrl.east |- koney.west) -- node [above, near start] {} (koney.west);
  \path [grayline] (ctrl.east |- tetragon.west) -- node [below, near start] {} (tetragon.west);

  \node (koney-down-right) at ($(koney.south east)+(-0.5cm,0)$) {};
  \node (ctrl-east-below) at ($(ctrl.east)+(-0.1cm,-0.2cm)$) {};
  \path [line, draw=ACMPurple] (koney-down-right) -- (koney-down-right |- ctrl-east-below)
  node [left, pos=0.35, name=path-4-label, scale=0.8] {\numExecThree{}} -- (ctrl-east-below);

  \node [anchor=west, name=tracing-policy-logo, left=of path-4-label, shift={(1.1cm,0)}] {\textcolor{ACMPurple}{\faFile*}};
  \node [below, anchor=east, align=right, name=tracing-policy-label, scale=0.7]
  at (tracing-policy-logo.west) {\ttfamily \textcolor{ACMPurple}{Tracing}\\\ttfamily \textcolor{ACMPurple}{Policy}};

  \node (koney-east-up) at ($(koney.east)+(0,0.1cm)$) {};
  \node (koney-east-down) at ($(koney.east)-(0,0.1cm)$) {};
  \node (koney-pod-mid) at ($(pod.north)+(-1.25cm,0.5cm)$) {};
  \path [line, draw=KubernetesBlue] (koney-east-up) 
  -- node [above, pos=0.73, scale=0.8, solid] {\numExecTwoAlt{}} (koney-east-up -| deployment.side 2);
  \path [line, draw=KubernetesBlue] (koney-east-down) -- (koney-pod-mid |- koney-east-down)
  -- (koney-pod-mid |- pod.side 2) -- node [above, pos=0.4, scale=0.8] {\numExecTwo{}} (pod.side 2);

  \path [line, draw=ACMPurple] (tetragon.north east) to[out=45,in=-60]
  node [right, midway, scale=0.8] {\numExecFour{}} (koney.south east);

  \path [grayline] (deployment.south)
  -- node [left, pos=0.4, scale=0.75, text=gray, shift={(-0.1cm,0)}] {creates} (pod.north);


  \path [grayline] (tetragon.east) -- node [below, midway, scale=0.75, text=gray, align=center] {installs\\eBPF probes}
  (pod.side 2 |- tetragon.east);

\end{tikzpicture}}}
  \caption{Deployment strategies for placing honeytokens.}
  \label{fig:exec-strategy}
\end{figure}

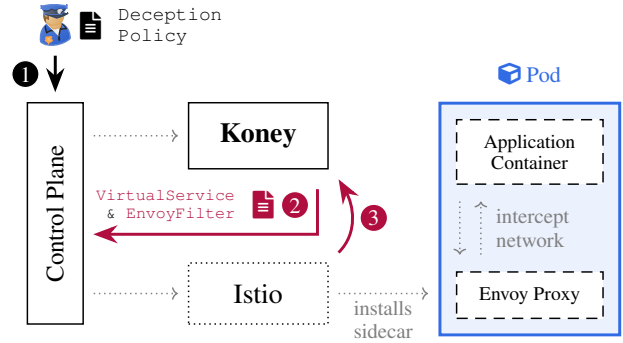
\begin{figure}[t]
  \centering
  \resizebox{\columnwidth}{!}{{\scalefont{0.75}\begin{tikzpicture}[scale=1.0,auto]
  \pgfdeclarelayer{background}
  \pgfsetlayers{background,main}

  \tikzstyle{block} = [rectangle, draw, node distance=0.6cm, outer sep=0.1cm, align=center]
  \tikzstyle{heptagon} = [regular polygon, draw, regular polygon sides=7, draw=none, fill=KubernetesBlue, text=white, outer sep=0.1cm]
  \tikzstyle{line} = [draw, -{Stealth[scale=1.0]}, thick]
  \tikzstyle{grayline} = [draw, densely dotted, -{>[scale=0.66]}, draw=gray, thin]

  \node [block, inner sep=0, minimum height=8em, minimum width=2em] (ctrl) {};
  \node [anchor=center, rotate=90, scale=0.9] at (ctrl.center) {Control Plane};

  \node [block, inner sep=0.2cm, minimum width=5em,
    right=of ctrl.north east, anchor=north west, shift={(0.2cm,0)}] (koney) {\bfseries\koney{}};
  \node [block, inner sep=0.2cm, minimum width=5em, densely dotted,
    right=of ctrl.south east, anchor=south west, shift={(0.2cm,0)}] (istio) {Istio};

  \node [block, densely dashed, inner sep=0.2cm, minimum width=7.5em, align=center, fill=white,
    anchor=north west, scale=0.7, shift={(1.5cm,-0.275cm)}] at (koney.north east) (app) {Application\\Container};
  \node [block, densely dashed, inner sep=0.2cm, minimum width=7.5em, node distance=0.7cm, fill=white,
    below=of app, scale=0.7] (sidecar) {Envoy Proxy};
  \begin{pgfonlayer}{background}
    \node [block, draw=KubernetesBlue, fill=KubernetesBlue!10, thick, fit=(app) (sidecar)] (pod) {};
    \node [above, text=KubernetesBlue, scale=0.8] at (pod.north) {\faCube{} Pod};
  \end{pgfonlayer}

  \node [police, scale=0.75, anchor=south, name=deception-policy-person, shift={(0,0.6cm)}] at (ctrl.north) {};
  \node [anchor=west, name=deception-policy-logo, right=of deception-policy-person, shift={(-1cm,0)}] {\faFile*};
  \node [below, anchor=west, align=left, name=deception-policy-label, scale=0.7]
  at (deception-policy-logo.east) {\ttfamily Deception\\\ttfamily Policy};

  \path [line] ($(deception-policy-person.south)+(0,-0.1cm)$) -- node [left, shift={(-0.07cm,0)}, scale=0.8] {\numIstioOne{}} (ctrl.north);
  \path [grayline] (ctrl.east |- koney.west) -- node [above, near start] {} (koney.west);
  \path [grayline] (ctrl.east |- istio.west) -- node [below, near start] {} (istio.west);

  \node (koney-down-right) at ($(koney.south east)+(-0.2cm,0)$) {};
  \node (ctrl-east-below) at ($(ctrl.east)+(-0.1cm,-0.2cm)$) {};
  \path [line, draw=ACMPurple!50!ACMRed] (koney-down-right) -- (koney-down-right |- ctrl-east-below)
  node [left, pos=0.35, name=path-4-label, scale=0.8] {\numIstioTwo{}} -- (ctrl-east-below);

  \node [anchor=west, name=tracing-policy-logo, left=of path-4-label, shift={(1.1cm,0)}] {\textcolor{ACMPurple!50!ACMRed}{\faFile*}};
  \node [below, anchor=east, align=right, name=tracing-policy-label, scale=0.6]
  at (tracing-policy-logo.west) {\ttfamily \textcolor{ACMPurple!50!ACMRed}{VirtualService}\\\ttfamily \& \textcolor{ACMPurple!50!ACMRed}{EnvoyFilter}};

  \node (app-south-left) at ($(app.south west)+(0.1cm,0)$) {};
  \node (app-south-right) at ($(app.south west)+(0.3cm,0)$) {};
  \node (sidecar-north-left) at ($(sidecar.north west)+(0.1cm,0)$) {};
  \node (sidecar-north-right) at ($(sidecar.north west)+(0.3cm,0)$) {};
  \path [grayline] (app-south-left) -- (sidecar-north-left);
  \path [grayline] (sidecar-north-right)
  -- node [right, midway, align=left, scale=0.75, shift={(0.1cm,0)}, text=gray, align=left]
  {intercept\\network} (app-south-right);

  \path [line, draw=ACMPurple!50!ACMRed] (istio.north east) to[out=45,in=-60]
  node [right, midway, scale=0.8] {\numIstioThree{}} (koney.south east);

  \node (sidecar-extra-west) at ($(sidecar.west)+(-0.2cm,0)$) {};
  \path [grayline] (istio.east) -- node [below, midway, scale=0.75, text=gray, align=center] {installs\\sidecar}
  (sidecar-extra-west |- istio.east);

\end{tikzpicture}}}
  \caption{Deployment strategies for placing HTTP-based traps.}
  \label{fig:istio-strategy}
\end{figure}

\koney{} automates deployment and monitoring of traps in Kubernetes
and is built using the Operator SDK%
~\cite{TheOperatorFrameworkAuthors:OperatorFramework},
a framework for developing Kubernetes operators with Golang.
Figures~\ref{fig:exec-strategy}~and~\ref{fig:istio-strategy}
show deployment strategies for honeytokens, and for HTTP-based traps, respectively.
\numExecOne{}~\koney{}'s reconciliation loop is invoked every time a
\textbf{\texttt{DeceptionPolicy}} resource is created, updated, or deleted. 
Updating a policy, e.g., to rotate traps,
is equivalent to deleting the old policy and creating a new one.



\subsection{Decoy Deployment Strategies}
\label{sec:operator:decoy-strategies}

The \koney{} operator can deploy decoys with different strategies.
The following sections describe the supported strategies
and their implementation in the \koney{} operator.

\subsubsection{\texttt{containerExec} Strategy}
\label{sec:operator:container-exec}

This strategy deploys file system honeytokens by executing shell commands directly in a container.
\numExecTwo{}~Commands are executed in the selected containers using the Kubernetes API,
by sending HTTP POST requests to the API server.
First, \koney{} creates the necessary directories in the container,
if they do not already exist, using the \texttt{mkdir} command.
Next, the operator creates honeytokens using the \texttt{echo} command,
redirecting the output to the desired file.
Note that we use \texttt{echo} instead of \texttt{touch} because in the case
where the file already includes content and we want to update it to be empty,
\texttt{touch} would not overwrite it.
To avoid command injection vulnerabilities,
the operator encodes the file's content in octal format
before executing the command in the container.
We do not use \texttt{base64} or hexadecimal encoding
because they might not be available in all containers and shells.
After creating the file, the operator checks that the file was created successfully
and that the content is correct using the \texttt{cat} command, verifying the output.
Finally, if the file is configured to be read-only in the deception policy,
the operator sets the file permissions to \texttt{444} using the \texttt{chmod} command.


\subsubsection{\texttt{volumeMount} Strategy}
\label{sec:operator:volume-mounts}

This strategy deploys honeytokens
by mounting a volume to the selected containers on the specified path.
The operator first creates a \texttt{Secret},
which is a Kubernetes object that contains sensitive data, with the content of the honeytoken.
\numExecTwoAlt{}~\koney{} then configures a volume in the selected deployment with
the \texttt{Secret} as the source and mounts it to the selected containers within the deployment.
Kubernetes will then re-create the pods with the mounted honeytokens
because we changed the deployment, which serves as a blueprint for creating pods.
The operator also sets the volume to be read-only, if specified in the policy,
to prevent the honeytoken from being modified from within the container.


\subsubsection{\texttt{istio} Decoy Strategy}
\label{sec:operator:istio-decoy}


This strategy first requires the Istio service mesh~\cite{TheIstioAuthors:Istio}
to be installed in the cluster.
A service mesh enables system operators to centrally monitor and control all cluster traffic.
Istio places a reverse proxy in front of application containers
by putting a sidecar container into every pod.
The sidecar intercepts all incoming and outgoing traffic
and applies the rules defined in Istio's \texttt{VirtualService} resources.
\numIstioTwo{}~\koney{} creates a virtual service
for each trap specified in the deception policy.
If the trap is a fixed HTTP response,
the operator generates a \texttt{HTTPDirectResponse} route action.
For traps that require modifying the HTTP response headers,
the operator generates a \texttt{HTTPRewrite} route action.
Finally, for traps that require modifying the body of the HTTP response,
we instead create an \texttt{EnvoyFilter} resource with a WebAssembly (WASM) extension.
Our WASM extension, a compiled \mbox{C++} program,
implements several ``engines'' for modifying HTTP bodies.

\subsection{Captor Deployment Strategies}
\label{sec:operator:captor-strategies}

Captors are responsible for monitoring access attempts to the decoys.
The following sections describe \koney{}'s supported strategies for deploying captors.

\subsubsection{\texttt{tetragon} Strategy}
\label{sec:operator:tetragon}

This strategy first requires the Tetragon operator to be installed in the cluster.
Tetragon~\cite{TheTetragonAuthors:Tetragon}
can trace system calls and monitor file access attempts with eBPF~\cite{Rice2022:WhatEBPF},
a Linux kernel technology.
\numExecThree{}~\koney{} creates a \texttt{TracingPolicy} resource
for each trap specified in the deception policy.
This policy monitors function calls related to file access.
Specifically, we monitor \texttt{security\_file\_permission} and \texttt{security\_mmap\_file}
kprobes with the path of the honeytoken as the argument.
\numExecFour{}~We configure the tracing policy to perform a request
to a designated webhook URL when it detects a file access attempt.
This URL is the address of a web server managed by \koney{}, which,
when it receives a request from Tetragon, reads Tetragon's logs,
identifies the trap that was accessed, and logs the incident.

\subsubsection{\texttt{istio} Captor Strategy}
\label{sec:operator:istio-captor}

Decoys placed with Istio~(\S\ref{sec:operator:istio-decoy})
must also be monitored with Istio.
\numIstioTwo{}~\koney{} creates an \texttt{EnvoyFilter}
for each trap specified in the deception policy.
\numIstioThree{}~The filter contains a WASM extension that will invoke
a webhook URL managed by \koney{} when the affected HTTP request is matched.
When \koney{} receives a request from Istio, 
it reads Istio's logs, identifies the triggered trap, and logs the incident.
This strategy is best applied on HTTP routes 
that are prone to attacks. 

\subsection{Auxiliary Functions}

The \koney{} operator provides additional features to assist system operators,
as briefly outlined below.

\textbf{Policy Validation.}
\koney{} validates deception policies before deploying traps.
This includes validating syntax 
and avoiding policy conflicts,
such as two policies attempting to place a honeytoken at the same file path.
However, \koney{} cannot guarantee that traps will not interfere with
the genuine flow of an application they target.


\textbf{Existing Resources.}
The \texttt{mutateExisting} field in a deception policy
specifies if traps should be deployed to already running workloads.
If enabled, \koney{} targets all matching workloads.
If disabled, traps are applied only to newly created resources post-policy creation.

\textbf{Workload Annotations.}
Every time \koney{} modifies a workload,
it places an annotation on that workload that contains
JSON-encoded metadata about the deployed traps.
Annotations are used by \koney{} during cleanup to identify what needs to be removed.

\textbf{Status Conditions.}
\koney{} tracks the deployment status of traps via so-called status conditions,
applied directly to deception policy objects.
These conditions report the number of deployed traps and indicate possible errors,
helping system operators troubleshoot deployment issues.

\textbf{Alerts.}
When a honeytoken is accessed or an HTTP-based trap is triggered,
\koney{} outputs a JSON-formatted log line to standard output in
the \texttt{alerts} container within the operator's pod.
We assume that system operators have monitoring software to centrally process these alerts.

\section{Evaluation}
\label{sec:evaluation}

We examine three questions:
(\S\ref{sec:evaluation:use-cases})
Which cyber deception techniques can be modeled in our policy documents and implemented by \koney{}?
(\S\ref{sec:evaluation:properties})
What are the trade-offs of \koney{}'s deployment strategies?
(\S\ref{sec:evaluation:performance})
How fast can \koney{} create, modify, and remove traps? 


\subsection{Use Case Coverage}
\label{sec:evaluation:use-cases}


Table~\ref{tab:evaluation} compares typical use cases
of application layer cyber deception~(\S\ref{sec:problem})
with a representative sample of related tools and frameworks,
which we also describe in related work~(\S\ref{sec:literature}).
This comparison is mainly dictated by the architectural constraints of specific frameworks.
\koney{} and Cloxy~\cite{Fraunholz2018:CloxyContextawareDeceptionasaService}
employ reverse proxies, theoretically allowing unrestricted modification of all web traffic.
Baitroute~\cite{Sen:Baitroute} is a software library integrated at compile-time,
designed to embed new deceptive endpoints without modifying existing ones.
HASH~\cite{DataDog:HASHHTTPAgnostic} is a classic honeypot deployed alongside applications,
which inhibits HASH from modifying the application traffic.

\begin{table}[th]
  \begin{threeparttable}
    \centering
    \caption{Use case coverage of application layer
      cyber deception tools and frameworks (as of April 2025)}
    \label{tab:evaluation}
    \newcommand\RHead[1]{\rotatebox{90}{\varwidth{\linewidth}#1\endvarwidth}}
\newcommand{\methodheader}[1]{{\RHead{#1}}} 

\newcommand{\EmptyMark}{{\color{black!10}$\circlet$}}
\newcommand{\HalfMark}{{\color{black!65}{$\circletfillhl$}}}
\newcommand{\FullMark}{$\circletfill$}

\newcommand{\NothingMark}{{\color{black!65}---}}

\newcommand{\YesMark}{\ding{52}}
\newcommand{\NoMark}{{\color{gray!30}\ding{56}}}
\newcommand{\UnknownMark}{\textbf{?}}

\setlength{\tabcolsep}{4.5pt}

\NewKeyValTable{Evaluation}{
  prop: align=X, head={~};
  Koney: align=c, head=\methodheader{\bfseries \koney{}};
  Cloxy: align=c, head=\methodheader{Cloxy~\cite{Fraunholz2018:CloxyContextawareDeceptionasaService}};
  Baitroute: align=c, head=\methodheader{Baitroute~\cite{Sen:Baitroute}};
  HASH: align=c, head=\methodheader{HASH~\cite{DataDog:HASHHTTPAgnostic}};
  DcyFS: align=c, head=\methodheader{DcyFS~\cite{Taylor2018:HiddenPlainSight}};
  Beesting: align=c, head=\methodheader{Beesting~\cite{Pichler:Beesting}};
}

\begin{KeyValTable}[%
    backend=tabularx,
    width=\columnwidth,
    shape=onepage,
    nobg=true,
    valign=c,
  ]{Evaluation}
  \Row{
    prop={Open-Source Framework},
    Koney=\YesMark{},
    Cloxy=\NoMark{},
    Baitroute=\YesMark{},
    HASH=\YesMark{},
    DcyFS=\NoMark{},
    Beesting=\YesMark{},
  }
  \Row{
    prop={Capable of Orchestration\tnote{1}},
    Koney=\YesMark{},
    Cloxy=\NoMark{},
    Baitroute=\NoMark{},
    HASH=\NoMark{},
    DcyFS=\YesMark{},
    Beesting=\NoMark{},
  }

  \MidRule

  \Row{
    prop={Honeytokens},
    Koney=\FullMark{}, 
    Cloxy=\NothingMark{},
    Baitroute=\NothingMark{},
    HASH=\NothingMark{},
    DcyFS=\FullMark{},
    Beesting=\FullMark{},
  }
  \Row{
    prop={Honeydocuments},
    Koney=\FullMark{}, 
    Cloxy=\NothingMark{},
    Baitroute=\NothingMark{},
    HASH=\NothingMark{},
    DcyFS=\FullMark{},
    Beesting=\EmptyMark{},
  }
  \Row{
    prop={Honeydirectories},
    Koney=\FullMark{}, 
    Cloxy=\NothingMark{},
    Baitroute=\NothingMark{},
    HASH=\NothingMark{},
    DcyFS=\FullMark{},
    Beesting=\EmptyMark{},
  }

  \MidRule

  \Row{prop={\bfseries Fixed HTTP Responses}}
  \Row{
    prop={URL Redirection},
    Koney=\FullMark{}, 
    Cloxy=\FullMark{}, 
    Baitroute=\FullMark{}, 
    HASH=\FullMark{}, 
    DcyFS=\NothingMark{},
    Beesting=\NothingMark{},
  }
  \Row{
    prop={Static Page, Honeyurl},
    Koney=\FullMark{}, 
    Cloxy=\FullMark{}, 
    Baitroute=\FullMark{}, 
    HASH=\FullMark{}, 
    DcyFS=\NothingMark{},
    Beesting=\NothingMark{},
  }
  \Row{
    prop={\enskip Upload Sinkhole},
    Koney=\hspace{0.425em}\HalfMark{}\tnote{2}, 
    Cloxy=\FullMark{}, 
    Baitroute=\hspace{0.425em}\HalfMark{}\tnote{2}, 
    HASH=\hspace{0.425em}\HalfMark{}\tnote{2}, 
    DcyFS=\NothingMark{},
    Beesting=\NothingMark{},
  }

  \Row[above=1ex]{prop={\bfseries HTTP Header Modification}}
  \Row{
    prop={Version Trickery, Fake Banner},
    Koney=\FullMark{}, 
    Cloxy=\FullMark{}, 
    Baitroute=\hspace{0.425em}\HalfMark{}\tnote{3}, 
    HASH=\EmptyMark{}, 
    DcyFS=\NothingMark{},
    Beesting=\NothingMark{},
  }
  \Row{
    prop={Status Code Tampering},
    Koney=\FullMark{}, 
    Cloxy=\hspace{0.425em}\HalfMark{}\tnote{4}, 
    Baitroute=\hspace{0.425em}\HalfMark{}\tnote{3}, 
    HASH=\EmptyMark{}, 
    DcyFS=\NothingMark{},
    Beesting=\NothingMark{},
  }
  \Row{
    prop={Cookie Tampering},
    Koney=\EmptyMark{}, 
    Cloxy=\hspace{0.425em}\HalfMark{}\tnote{5}, 
    Baitroute=\EmptyMark{}, 
    HASH=\EmptyMark{}, 
    DcyFS=\NothingMark{},
    Beesting=\NothingMark{},
  }

  \Row[above=1ex]{prop={\bfseries HTTP Body Modification}}
  \Row{
    prop={robots.txt Mod., Disallow Inj.},
    Koney=\FullMark{}, 
    Cloxy=\FullMark{}, 
    Baitroute=\EmptyMark{}, 
    HASH=\EmptyMark{}, 
    DcyFS=\NothingMark{},
    Beesting=\NothingMark{},
  }
  \Row{
    prop={HTML, CSS, JS Modification},
    Koney=\FullMark{}, 
    Cloxy=\FullMark{}, 
    Baitroute=\EmptyMark{}, 
    HASH=\EmptyMark{}, 
    DcyFS=\NothingMark{},
    Beesting=\NothingMark{},
  }
  \Row{
    prop={\enskip Tracking Links},
    Koney=\FullMark{}, 
    Cloxy=\FullMark{}, 
    Baitroute=\EmptyMark{}, 
    HASH=\EmptyMark{}, 
    DcyFS=\NothingMark{},
    Beesting=\NothingMark{},
  }
  \Row{
    prop={\enskip Code Obfuscation},
    Koney=\EmptyMark{}, 
    Cloxy=\FullMark{}, 
    Baitroute=\EmptyMark{}, 
    HASH=\EmptyMark{}, 
    DcyFS=\NothingMark{},
    Beesting=\NothingMark{},
  }
  \Row{
    prop={\enskip Weak Code Injection},
    Koney=\FullMark{}, 
    Cloxy=\FullMark{}, 
    Baitroute=\EmptyMark{}, 
    HASH=\EmptyMark{}, 
    DcyFS=\NothingMark{},
    Beesting=\NothingMark{},
  }
  \Row{
    prop={\enskip GET / POST Param. Injection},
    Koney=\FullMark{}, 
    Cloxy=\FullMark{}, 
    Baitroute=\EmptyMark{}, 
    HASH=\EmptyMark{}, 
    DcyFS=\NothingMark{},
    Beesting=\NothingMark{},
  }
  \Row{
    prop={\enskip\enskip Hidden Form Fields},
    Koney=\FullMark{}, 
    Cloxy=\FullMark{}, 
    Baitroute=\EmptyMark{}, 
    HASH=\EmptyMark{}, 
    DcyFS=\NothingMark{},
    Beesting=\NothingMark{},
  }
\end{KeyValTable}

    \begin{tablenotes}
      \footnotesize
      \item[1] Does this framework help manage and coordinate
      the deployment of several traps across multiple containers or processes?
      \item[2] No native support, but can be achieved by redirecting to a honeypot.
      \item[3] Baitroute would replace the entire HTTP response with a new one.
      \item[4] Designed to deceive automated web scanners, not configurable.
      \item[5] Designed to append decoy cookies, not to modify existing ones.
    \end{tablenotes}
  \end{threeparttable}
\end{table}

This comparison omits works such as
Decepto~\cite{Santoro2024:DemoCloudnativeCyber},
HoneyKube%
~\cite{
  Gupta2023:HoneyKubeDesigningDeploying,
  Gupta2021:HoneyKubeDesigningHoneypot},
or the Kubernetes Storm Center~\cite{TheKubernetesStormCenterAuthors:KubernetesStormCenter},
as they use Kubernetes to build complete
``honeyclusters'' but do not inject traps into the application layer.

\subsection{Operational Trade-Offs}
\label{sec:evaluation:properties}

Kahlhofer and Rass 
introduced various properties to evaluate technical methods that implement deception techniques%
~\cite{Kahlhofer2024:ApplicationLayerCyber}.
We focus on evaluating a subset of them, namely
detectability (how well a method can detect attacks),
simplicity, maintainability, scalability,
inconspicuousness (how hard it is for attackers to identify a technical method),
and non-interference with genuine application assets,
or as Gupta~et~al. put it, 
to ``not make the [overall] system less secure''%
~\cite{Gupta2021:HoneyKubeDesigningHoneypot,Gupta2023:HoneyKubeDesigningDeploying}.

\textbf{The Operator Pattern.}
Separating policy documents from the tools that implement them
improves maintainability and scalability, at the cost of increased complexity.
A Kubernetes operator is easy to integrate into a cluster, similar to a plugin;
however, since this also grants them broad privileges,
it risks disrupting existing workloads.
The Kubernetes Operator Threat Matrix~\cite{ControlPlane:KubernetesOperatorThreat}
is a useful resource to assess the risks of operators, in general.

\textbf{Placing Honeytokens in Containers.}
\koney{} places honeytokens by executing shell commands
in running containers~(\S\ref{sec:operator:container-exec})
or by defining volume mounts~(\S\ref{sec:operator:volume-mounts}).

Executing shell commands is simple, flexible,
leaves no trace for attackers,
and can be done without restarting applications.
This approach relies on a file system that is not read-only
and binaries such as \texttt{sh} and \texttt{echo},
which makes it difficult to support every container
-- impossible even for distroless container images,
as they only include the application and necessary runtime dependencies.
Also, it is generally not good practice to execute shell commands in production containers%
~(\S\ref{sec:discussion}).

Defining volume mounts solves these issues, while also improving transparency,
since system operators can easily locate them in manifests.
However, this requires restarting the application
because volume mounts cannot be attached to running containers.

\textbf{Monitoring with eBPF.}
\koney{} uses eBPF to detect when a process accesses a honeytoken~(\S\ref{sec:operator:tetragon}).
Since eBPF programs run in kernel space,
attackers can hardly bypass or identify this detection mechanism.
eBPF is a well-established technology within the Kubernetes ecosystem, 
thus providing good maintenance and scalability characteristics.
All eBPF programs operate with a limited instruction set
and have read-only access to kernel data structures,
ensuring workloads remain undisturbed.

\textbf{Reverse Proxies.}
\koney{} relies on the Istio service mesh to place
Envoy proxies in front of application containers~(\S\ref{sec:operator:istio-decoy}, \S\ref{sec:operator:istio-captor}).
A service mesh, while not necessarily easy to set up initially,
provides a very scalable and typically well-maintained foundation~\cite{Li2019:ServiceMeshChallenges}.
Reverse proxies can manage and monitor all incoming and outgoing application traffic,
allowing them to detect attacks on traps
and potentially also taking on additional intrusion detection functions.
Reverse proxies often come with a significant latency overhead%
~\cite{Zhu2023:DissectingOverheadsService},
which could make them unsuitable for high-performance scenarios.
These interposed systems, if not properly configured,
may interfere with application workloads
or reveal their presence to attackers
if proxies append identifiable details.

\subsection{Operational Performance}
\label{sec:evaluation:performance}

\koney{} can be added to any Kubernetes cluster
in minutes by executing a few \texttt{kubectl} commands.
\koney{} deploys traps immediately after a new \texttt{DeceptionPolicy} resource
is added to the cluster.
Most traps are deployed within seconds; some require a container restart,
resulting in a short downtime unless replicas are available.
Alerts that appear when traps are triggered are slightly delayed
because Tetragon and Istio logs are currently processed asynchronously
every 30 seconds.
We did not conduct a quantitative evaluation of these processes
because it is evident that automated workflows are
inherently quicker than the manual deployment of cyber deception techniques.

\section{Discussion}
\label{sec:discussion}

This section discusses our design choices 
and identifies challenges and opportunities for future work.

\textbf{Treating deception technology as policy objects, like Kubernetes does,
  makes it manageable and accelerates its adoption.}
Deception technologies frequently demand unconventional tricks to maintain secrecy, 
which system operators are reluctant to adopt.
A first useful step is to increase transparency for system operators
by making deception technologies manageable policy objects.
We presented \koney{} to 6~security engineers in our team, who
appreciated that \koney{} annotates workloads that it manipulated, 
that policy documents report their deployment status,
and that each trap's deployment is automatically validated
by programmatically accessing it after creation.

\textbf{When adopting deception technology, system operators
  must choose between having many dynamic parts in production, risking interference,
  or integrating deception technology earlier in the software development life cycle.}
Authors of policy documents still need to be careful
not to impair or break existing workloads with their traps.
Mounting honeytokens via bind mounts or executing OCI hooks as Beesting%
~\cite{Pichler:Beesting,Pichler2025:BeestingCantTouch}
does can be done with minimal interference,
while intercepting HTTP communication with reverse proxies is riskier.
In addition, \koney{}'s deployment strategies may also trigger other security tools.
During tests in a AWS EKS cluster, 
AWS GuardDuty generated security alerts due to suspicious activities by privileged pods
because our ``exec'' strategy executes shell scripts in pods (\S\ref{sec:operator:container-exec}),
which is also typical for malware.
Although allow-listing \koney{} resolves this, ideally,
deception technology would be integrated earlier in the software development life cycle,
e.g., at compile-time like Baitroute~\cite{Sen:Baitroute}.
However, this is often challenging or impossible, especially with third-party software.
Consequently, we encourage more research on the risk levels posed by cyber deception technologies.

\textbf{Service meshes and reverse proxies are flexible and extensible,
  but research is needed to resolve their performance overhead.}
Although \koney{} and Cloxy~\cite{Fraunholz2018:CloxyContextawareDeceptionasaService}
do not yet support certain use cases of deception in web applications (\S\ref{sec:evaluation}),
their flexible proxy-based architecture should allow future extensions.
However, the costly setup process for a service mesh,
including Istio~\cite{TheIstioAuthors:Istio} and Linkerd~\cite{TheLinkerdAuthors:Linkerd},
and its invasive nature remain disadvantageous.
Performance also deteriorates,
as reverse proxies consume extra memory and CPU resources and increase network latency%
~\cite{Zhu2023:DissectingOverheadsService}.
Still, their prevalence in research~\citeReverseProxySolutions{}
indicates a lack of practical alternatives.
There are function-hooking-based alternatives
that are also more resource-efficient%
~\cite{Kahlhofer2024:ApplicationLayerCyber},
but they add a considerable amount of system complexity.

\textbf{eBPF is a great choice for designing captors for containerized workloads,
  but its peculiarities should not be underestimated.}
eBPF is a popular and natural fit for cloud-native environments%
\cite{Soldani2023:EBPFNewApproach}.
Monitoring honeytoken access attempts with Tetragon (\S\ref{sec:operator:tetragon})
is simple, as Tetragon provides \texttt{TracingPolicy} custom resources
and a guide to file monitoring~\cite{Kourtis2024:FileMonitoringEBPF}.
Falco~\cite{TheFalcoAuthors:Falco} offers a similar solution,
which we plan to support in future work.
Although these tools exist, developing eBPF programs remains challenging.
We experienced some issues when installing eBPF hooks,
likely due to incompatibilities with Linux kernel versions in our test clusters.
Additionally, Tetragon throttles event generation to avoid overloading the kernel,
potentially dropping honeytoken access events in busy clusters.
Nonetheless, eBPF technology, if engineered with care, appears favorable over alternatives
such as custom file systems~\cite{Taylor2018:HiddenPlainSight}
or function hooking~\cite{Kern2024:InjectingSharedLibraries}. 

\textbf{Cyber deception policy documents provide structure
  to the intricate nature of deception technology.}
The YAML structure proposed in this work (\S\ref{sec:policies}) is a first step
to provide a concrete structure for deception technology.
Good API design is an iterative process between users and developers.
For the next iteration, we would consider whether deployment strategies for decoys and captors
should rather be placed elsewhere 
to further separate the roles of deception designers and technical engineers.
Our current structure also lacks the means to write conditional and stateful deception policies,
such as activating deception objects only for unauthenticated users
or after detecting anomalous behavior patterns.

\textbf{Minimizing the operational cost
  of deploying deception technology ultimately benefits defenders.}
Our core idea is to make the use of deception technology
as straightforward as flipping a feature flag.
This is not only practical, but also intriguing in theory.
Operational costs can be an inhibitor to effective moving target defense
(e.g., leveraging game theory or other methods)
if the defender shall adapt deception strategies dynamically.
This problem has been observed theoretically
and findings indicate that neglecting the costs of switching strategies may explain
why individuals behave differently from what
game theory predicts for a rational adversary and defender.%
~\cite{Rass2017:CostGamePlaying,Rass2018:PasswordSecurityGame}.

\subsection{Extensions}
\label{sec:discussion:extensions}

Since \koney{} operates directly inside the cluster with elevated privileges,
its capabilities can be easily extended to achieve
any cyber deception technology that can be installed at runtime.
Of the 19 technical methods described by
Kahlhofer and Rass~\cite{Kahlhofer2024:ApplicationLayerCyber}, 
the current design of \koney{} could support 17 of them.
This includes adding new pods, containers, or services to the cluster,
running shell commands when containers start, modifying environment variables,
and even installing custom file systems such as DcyFS%
~\cite{Taylor2018:HiddenPlainSight}
if \koney{} is installed with privileged access to nodes.
Only techniques that require access to source code,
modification of container images, the build pipeline,
or similar ``early'' parts of the software development life cycle are not addressed by \koney{}.

\koney{} could easily be extended with policies that deploy classic honeypots%
~\cite{Nawrocki2016:SurveyHoneypotSoftware}
for protocols such as FTP, SSH, or SMTP
as long as container images are available for them; similar to the \mbox{T-Pot} project%
~\cite{DeutscheTelekomSecurityGmbH:TPot},
which facilitates the deployment of more than 20 different containerized honeypots.
Deception techniques for domain-specific applications
are also realizable if they can be installed by reconfiguring application artifacts.
Adding new configuration files to containers can be done with
the ``exec'' method (\S\ref{sec:operator:container-exec})
or with volume mounts (\S\ref{sec:operator:volume-mounts}).
Similarly, \koney{} could also change or add environment variables in manifests.
Deceptive data can also be inserted into databases if
\koney{} connects to them within the cluster using provided database credentials.

\section{Related Work}
\label{sec:literature}

The evolution of cyber deception has advanced from honeypots%
~\cite{
  Spitzner2003:HoneypotsCatchingInsider,
  Provos2004:VirtualHoneypotFramework}
to honeytokens%
~\cite{Spitzner2003:HoneytokensOtherHoneypot},
and is currently focused on at least three problems:
finding effective traps%
~\cite{%
  Kahlhofer2024:HoneyquestRapidlyMeasuring,
  Sahin2022:ApproachGenerateRealistic,
  Sahin2022:MeasuringDevelopersWeb,
  Sahin2020:LessonsLearnedSunDEW,
  Ferguson-Walter2019:TularosaStudyExperimental,
  Ferguson-Walter2021:ExaminingEfficacyDecoybased,
  Bercovitch2011:HoneyGenAutomatedHoneytokens,
  Bowen2010:AutomatingInjectionBelievable},
recently aided by generative AI%
~\citeGenerativeHoneypots{};
developing game-theoretical models%
~\cite{
  Zhu2021:SurveyDefensiveDeception,
  Pawlick2019:GametheoreticTaxonomySurvey};
and creating novel cyber deception techniques%
~\cite{%
  Han2018:DeceptionTechniquesComputer,
  Zhang2021:ThreeDecadesDeception,
  Fraunholz2018:DemystifyingDeceptionTechnology,
  Fan2018:EnablingAnatomicView}.

The literature on the operational aspects of modern cyber deception 
is relatively scarce%
~\cite{
  Kahlhofer2024:ApplicationLayerCyber,
  Fraunholz2017:DeploymentStrategiesDeception},
possibly due to the assumption that the industry could address these challenges independently.
However, the limited adoption of these technologies suggests underlying research-worthy issues.
Although many works built proof-of-concepts 
to demonstrate novel cyber deception techniques%
~\cite{
  Araujo2014:PatchesHoneyPatchesLightweight,
  Han2017:EvaluationDeceptionBasedWeb,
  Gavrilis2007:FlashCrowdDetection,
  Fraunholz2018:DefendingWebServers,
  Kern2024:InjectingSharedLibraries,
  Petrunic2015:HoneytokensActiveDefense,
  BenSalem2011:DecoyDocumentDeployment,
  Voris2015:FoxTrapThwarting,
  Yuill2004:HoneyfilesDeceptiveFiles,
  Angeli2024:FalseFlavorHoneypot,
  Barron2021:ClickThisNot,
  Reti2023:SCANTRAPProtectingContent}
and the deployment of classic honeypot software is well-surveyed and reviewed%
~\cite{
  Han2018:DeceptionTechniquesComputer,
  Fraunholz2017:DeploymentStrategiesDeception,
  Nawrocki2016:SurveyHoneypotSoftware},
we have not found any publications that present an
open-source cyber deception orchestration framework for the application layer.
It is also crucial to note that a substantial portion of the literature
uses the term ``framework'' to refer to theoretical frameworks
rather than software frameworks and tools.

\subsection{Works on Deception in File Systems}

Beesting%
~\cite{Pichler2025:BeestingCantTouch,Pichler:Beesting}
places honeytokens in Kubernetes with OCI hooks and, like \koney{}, volume mounts.
An OCI hook is code that runs on container events; they use
a hook on container startup to create honeytokens in the container.
Unlike \koney{}, which mutates manifests via the Kubernetes API,
Beesting uses the node resource interface (NRI)%
~\cite{ThecontainerdAuthors:NRINodeResource},
a standardized method for adding custom logic to container runtimes.
For monitoring file access, Beesting also uses eBPF programs, but natively, without Tetragon.

DcyFS%
~\cite{Taylor2018:HiddenPlainSight}
is an overlay file system that transparently injects honeytokens. 
Unique to this approach is that these file system views are created on a per-process basis only.

Many early works implemented tools for placing honeytokens%
~\cite{
  Voris2015:FoxTrapThwarting,
  BenSalem2011:DecoyDocumentDeployment,
  Yuill2004:HoneyfilesDeceptiveFiles,
  Bowen2009:BaitingAttackersUsing
},
but these were typically meant to assist with experimental evaluations
and were not described in detail or made publicly available.

\subsection{Works on Deception in Web Applications}

Cloxy%
~\cite{Fraunholz2018:CloxyContextawareDeceptionasaService}
is a deception-as-a-service software framework that implements most
HTTP-based traps discussed herein~(\S\ref{sec:problem}), much like \koney{}.
Cloxy uses mitmproxy%
~\cite{Cortesi:MitmproxyFreeOpen}
as a reverse proxy placed in front of applications to intercept network communication.
\koney{} uses Istio, which utilizes the Envoy proxy%
~\cite{TheEnvoyProjectAuthors:Envoy}.
Envoy is built for real-world use, offering much better performance%
~\cite{Elkhatib2023:EvaluationServiceMesh}
than mitmproxy, which is intended for rapid prototyping. 

Most works use reverse proxies to add deception to HTTP communication%
~\citeReverseProxySolutions{}.
If access to the source code and recompiling applications is possible,
which \koney{} does not assume, then software libraries can be used:
Baitroute%
~\cite{Sen:Baitroute}
is a library for Go, Python, and JavaScript that can serve vulnerable-looking endpoints.
Its rules define what traps are served and closely resemble
the structure of \koney{}'s definition of traps for fixed HTTP responses%
~(\S\ref{sec:policies:web}).

Exceptions to approaches based on reverse proxies include the work of
Kern~\cite{Kern2024:InjectingSharedLibraries}, 
who used \texttt{LD\_PRELOAD} to intercept libc functions responsible for network communication,
and the work of
Reti~et~al.~\cite{Reti2023:HoneyInfiltratorInjecting}, 
who modified packages directly with Netfilter and Scapy%
~\cite{Biondi:Scapy}.

In the context of Kubernetes, HoneyKube%
~\cite{
  Gupta2023:HoneyKubeDesigningDeploying,
  Gupta2021:HoneyKubeDesigningHoneypot}
was among the first works to use Kubernetes as a platform to deploy honeypot microservices.
The Kubernetes Storm Center~\cite{TheKubernetesStormCenterAuthors:KubernetesStormCenter}
aims to gather threat intelligence by building complete ``honeyclusters''.
Decepto~\cite{Santoro2024:DemoCloudnativeCyber}
generates decoys as clones of production microservices running in Kubernetes.
All of these studies share a focus on classic network-based honeypots,
and did not inject traps into the application layer.
KubeDeceive~\cite{Aly2023:KubeDeceiveUnveilingDeceptive}
is different again, as they intercept API calls to the control plane.

\subsection{Works on Deception Policies and Operators}

Although abstract concepts, processes, and strategies for cyber deception have been proposed%
~\cite{
  Zhang2021:ThreeDecadesDeception,
  DeFaveri2016:DesigningAdaptiveDeception,
  DeFaveri2016:GoalDrivenDeceptionTactics},
we have not yet seen any reports on how to describe these with code. 
The Honeyquest tool%
\cite{Kahlhofer2024:HoneyquestRapidlyMeasuring},
intended to measure the enticingness of cyber deception,
introduced a minimal language called \mbox{HoneYAML} to describe their cyber traps.
Baitroute%
~\cite{Sen:Baitroute}
and HASH%
~\cite{DataDog:HASHHTTPAgnostic},
two tools for creating low-interaction web-based honeypots,
introduced a YAML structure to describe their traps,
similar to the specification of HTTP-based traps
in our policy documents~(\S\ref{sec:policies:web}).

Using Kubernetes operators for policy enforcement is a common practice.
The Open Policy Agent~(OPA)%
~\cite{TheOpenPolicyAgentAuthors:OpenPolicyAgent}
allows system operators to write policy documents as code and
moves the policy decision-making process out of the software and into OPA.
Kyverno%
~\cite{TheKyvernoAuthors:Kyverno}
supports security policies that
can validate, mutate, generate, and clean up any Kubernetes resource.
However, a general-purpose Kubernetes operator
for cyber deception has not yet appeared in the literature.


\section{Conclusion}
\label{sec:conclusion}

This work conceptualized cyber deception policies,
which describe deception technology ``as code'',
and \koney{}, a cyber deception orchestration framework.
Kubernetes proved to be an ideal enabling technology for \koney{},
allowing versatile integration of cyber deception.
The prevalence of Kubernetes in industry also brings synergies, 
as research and application share the same platform.
We demonstrated that treating deception technology as policy objects
makes it manageable for system operators
and helps to separate responsibilities between deception technique authors,
software application developers, and system operators.
\koney{}'s technical implementation employs cloud-native technologies, 
such as services meshes and eBPF, which we also see as beneficial
for future research on cyber deception, beyond use cases for the application layer,
and for platforms other than Kubernetes.

\ifdefined\AnonymousReview\else
  \section*{Acknowledgments}
  We thank Farooq, Anis, Josef, Werner, and the anonymous reviewers
  for their useful feedback. 
  We acknowledge the support of the MUR PNRR project
  PE SERICS \textendash{} SecCO (PE00000014) CUP D33C22001300002
  funded by the European Union under NextGenerationEU.
\fi

\clearpage

\IEEEtriggeratref{77}
\IEEEtriggercmd{\enlargethispage{-5.35in}}

\bibliographystyle{IEEEtranS}
{\small\sloppy\hfuzz=2pt\hbadness=2500\bibliography{abbrev,references}}

\clearpage

\appendices
\section{Sample Cyber Deception Policies}
\label{sec:appendix:examples}

This section presents \texttt{DeceptionPolicy} samples
that can be readily deployed with \koney{}.

\subsection{Samples for Traps in File Systems}
\label{sec:appendix:examples:fs}

Listing~\ref{lst:appendix:honeyfiles} shows a sample cyber deception policy
for placing a honeytoken in all workloads
with the \texttt{op/honeytoken=true} label
and a honeydocument in all workloads
with the \texttt{op/honeydocument=true} label,
but only in containers whose name starts with \mbox{``app-''}.
Both files are deployed by executing shell commands in the container.
The document is downloaded from a specified URL.
File monitoring is done with Tetragon.

\begin{lstlisting}[
  language=yaml,
  label=lst:appendix:honeyfiles,
  caption={Samply policy for placing honeyfiles}
]
# please refer to the latest api version at
# (*@\YAMLcommentstyle\url{\repository}@*)
---
apiVersion: ...
kind: DeceptionPolicy
metadata:
  name: sample-honeyfiles
spec:
  strictValidation: true
  mutateExisting: true

  traps:
    - filesystemHoneytoken:
        path: /run/secrets/service_token
        content: very-secret-token
        readOnly: true

      match:
        any:
          - resources:
              containerSelector: "*"
              selector:
                matchLabels:
                  op/honeytoken: true

      decoyDeployment:
        strategy: containerExec
      captorDeployment:
        strategy: tetragon

    - filesystemHoneydocument:
        path: /root/passwords.docx
        source: https://srv.test/honey.docx
        readOnly: false

      match:
        any:
          - resources:
              containerSelector: "app-*"
              selector:
                matchLabels:
                  op/honeydocument: true

      decoyDeployment:
        strategy: containerExec
      captorDeployment:
        strategy: tetragon
\end{lstlisting}

Listing~\ref{lst:appendix:honeydirectories} shows a sample cyber deception policy
for setting up a honeydirectory in all workloads
with the \texttt{op/honeydirectory=true} label.
The directory will contain two files,
using the same specification as when placing honeytokens.
All files will be mounted as a volume inside all containers.
File monitoring is done with Tetragon.
Only resources that are created after the deception policy
was created will get this trap because \texttt{mutateExisting} is not set here.

\begin{lstlisting}[
  language=yaml,
  label=lst:appendix:honeydirectories,
  caption={Policy for setting up a honeydirectory}
]
# please refer to the latest api version at
# (*@\YAMLcommentstyle\url{\repository}@*)
---
apiVersion: ...
kind: DeceptionPolicy
metadata:
  name: sample-honeydirectory
spec:
  strictValidation: true
  mutateExisting: false

  traps:
    - filesystemHoneydirectory:
        path: /tmp
        content:
          - filesystemHoneytoken:
              path: auth_token
              content: very-secret-token
          - filesystemHoneytoken:
              path: config.ini
              content: ENVIRONMENT=prod

      match:
        any:
          - resources:
              selector:
                matchLabels:
                  op/honeydirectory: true

      decoyDeployment:
        strategy: volumeMount
      captorDeployment:
        strategy: tetragon
\end{lstlisting}

\subsection{Samples for Traps in Web Application}
\label{sec:appendix:examples:web}

Listing~\ref{lst:appendix:honeypot-redirect} shows a sample cyber deception policy
for setting up a HTTP redirect to a honeypot in all workloads
with the \texttt{op/honeypot-redirect=true} label.
Communication on all ports and in all containers will be affected.
The route is placed by utilizing Istio's reverse proxies.
Monitoring is also done with Istio.

\begin{lstlisting}[
  language=yaml,
  label=lst:appendix:honeypot-redirect,
  caption={Policy for setting up a redirect to a honeypot}
]
# please refer to the latest api version at
# (*@\YAMLcommentstyle\url{\repository}@*)
---
apiVersion: ...
kind: DeceptionPolicy
metadata:
  name: sample-honeypot-redirect
spec:
  strictValidation: true
  mutateExisting: true

  traps:
    - httpResponse:
        request:
          match: ^/admin$  # exact match
          method: GET      # GET only
        response:
          status: 302      # temp redirect
          headers:
            Location: http://honeypot.test

      match:
        any:
          - resources:
              ports: null  # all ports
              selector:
                matchLabels:
                  op/honeypot-redirect: true

      decoyDeployment:
        strategy: istio
      captorDeployment:
        strategy: istio
\end{lstlisting}

Listing~\ref{lst:appendix:aws-credentials} shows a sample cyber deception policy
for adding a new ``Disallow'' entry at the end of the ``robots.txt'' file,
when delivered over HTTP, in all workloads
with the \texttt{op/disallow-injection=true} label,
and for setting up a fixed HTTP response in all workloads
with the \texttt{op/aws-credentials=true} label.
The fixed HTTP response will only affect communication on port 80 and 8080.
The new route and the HTTP body modification is done by utilizing Istio's reverse proxies.
Monitoring is also done with Istio,
but only for the fixed HTTP response, not for the ``Disallow'' entry.

\raggedbottom

\begin{lstlisting}[
  language=yaml,
  label=lst:appendix:aws-credentials,
  caption={Policy for exposing fake credentials}
]
# please refer to the latest api version at
# (*@\YAMLcommentstyle\url{\repository}@*)
---
apiVersion: ...
kind: DeceptionPolicy
metadata:
  name: sample-aws-credentials
spec:
  strictValidation: true
  mutateExisting: true

  traps:
    - httpBodyMutation:
        request:
          match: ^/robots.txt$  # exact match
          method: GET           # GET only
        response:
          matchHeaders:
            Content-Type: text/html  # HTML only
          bodyMutations:
            - engine: regex
              match: >-
                (*@\YAMLvaluestyle (?s)(.*) @*)
              replace: >-
                (*@\YAMLvaluestyle \$1\textbackslash{}nDisallow: /.aws/credentials @*)

      match:
        any:
          - resources:
              selector:
                matchLabels:
                  op/disallow-injection: true

      decoyDeployment:
        strategy: istio
      captorDeployment:
        strategy: null

    - httpResponse:
        request:
          match: /.aws/credentials$  # ends with
          method: GET                # GET only
        response:
          status: 200
          headers:
            Content-Type: text/plain
          body: |
            (*@\YAMLvaluestyle [default] @*)
            (*@\YAMLvaluestyle aws\_access\_key\_id = ASLPVFCMPNXCFEX @*)
            (*@\YAMLvaluestyle aws\_secret\_access\_key = h8aQcFM64 @*)
            (*@\YAMLvaluestyle region = us-east-1 @*)

      match:
        any:
          - resources:
              ports:
                - 80
                - 8080
              selector:
                matchLabels:
                  op/aws-credentials: true

      decoyDeployment:
        strategy: istio
      captorDeployment:
        strategy: istio
\end{lstlisting}

Listing~\ref{lst:appendix:fake-banner} shows a sample cyber deception policy
for adding headers to HTTP responses in all workloads
with the \texttt{op/fake-banner=true} label.
Communication on all ports and in all containers will be affected.
The headers are added by utilizing Istio's reverse proxies.
No monitoring is deployed for this policy.

\begin{lstlisting}[
  language=yaml,
  label=lst:appendix:fake-banner,
  caption={Policy for adding HTTP headers to responses}
]
# please refer to the latest api version at
# (*@\YAMLcommentstyle\url{\repository}@*)
---
apiVersion: ...
kind: DeceptionPolicy
metadata:
  name: sample-fake-banner
spec:
  strictValidation: true
  mutateExisting: true

  traps:
    - httpHeaderMutation:
        request:
          match: "*"   # match all
          method: GET  # GET only
        response:
          setHeaders:
            Server: nginx/1.2.4
            X-ApiServer: honeypot.test

      match:
        any:
          - resources:
              ports: null  # all ports
              selector:
                matchLabels:
                  op/fake-banner: true

      decoyDeployment:
        strategy: istio
      captorDeployment:
        strategy: null
\end{lstlisting}

\end{document}